# Coexisting charge and magnetic orders in the dimer-chain iridate $Ba_5AlIr_2O_{11}$

J. Terzic[1], J. C. Wang[1,2,3], Feng Ye[1,2], W. H. Song[1,4], S. J. Yuan[1], S. Aswartham[1], L. E. DeLong[1], S.V. Streltsov[5,6], D.I. Khomskii[7] and G. Cao[1*]

[1]Center for Advanced Materials, Department of Physics and Astronomy, University of Kentucky, Lexington, KY 40506, USA

[2]Quantum Condensed Matter Division, Oak Ridge National Laboratory, Oak Ridge, Tennessee 37831, USA

[3]Department of Physics, Renmin University of China, Beijing, China

[4]Institute of Solid State Physics, Chinese Academy of Sciences, Hefei, China

[5]Institute of Metal Physics, 620041 Ekaterinburg, Russia

[6]Ural Federal University, 620002 Ekaterinburg, Russia

[7]II.Physikalisches Institut, Universitaet zu Koeln, Germany

*Abstract* We have synthesized and studied single-crystal $Ba_5AlIr_2O_{11}$ that features dimer chains of two inequivalent octahedra occupied by tetravalent $Ir^{4+}$($5d^5$) and pentavalent $Ir^{5+}$($5d^4$) ions, respectively. $Ba_5AlIr_2O_{11}$ is a Mott insulator that undergoes a subtle structural phase transition near $T_S = 210$ K and a magnetic transition at $T_M = 4.5$ K; the latter transition is surprisingly resistant to applied magnetic fields $\mu_o H \leq 12$ T, but sensitive to modest applied pressure ($dT_M/dp \approx +0.61$ K/GPa). All results indicate that the phase transition at $T_S$ signals an enhanced charge order that induces electrical dipoles and strong dielectric response near $T_S$. It is clear that the strong covalency and spin-orbit interaction (SOI) suppress double exchange in Ir dimers and stabilize a novel magnetic state that is neither S = 3/2 nor J = ½, but rather lies in an "intermediate" regime between these two states. The novel behavior of $Ba_5AlIr_2O_{11}$ therefore provides unique insights into the physics of SOI along with strong covalency in competition with double exchange interactions of comparable strength.

*Introduction* The 5d-electron based iridates are distinguished by strong spin-orbit interaction (SOI ~ 0.4 eV) that is comparable to the on-site Coulomb interaction (U ~ 0.5 eV), crystalline electric field interactions and Hund's rule coupling ($J_H$ ~ 0.5 eV). This unique circumstance creates a *delicate* balance between interactions that generate novel magnetic states and dielectric behaviors. A profound manifestation of this competition is the "$J_{eff}$ = 1/2 insulating state" first observed in $Sr_2IrO_4$ **[1-4]**. Indeed, in contrast to conventional wisdom, the iridates are much more likely to host a magnetic, insulating ground state with a very small magnetic moment **[4-8]** that is more strongly coupled to the lattice rather than other degrees of freedom **[2, 9-16]**.

Most studies of the iridates have focused on the tetravalent iridates since the $Ir^{4+}$($5d^5$) ion provides four d-electrons to fill the lower $J_{eff}$ = 3/2 bands, and one electron to partially fill the upper $J_{eff}$ = 1/2 band that lies closest to the Fermi energy, and therefore dominates underlying physical properties. Very limited studies have been devoted to iridates having pentavalent $Ir^{5+}$($5d^4$) ions in part, because the strong SOI limit is expected to impose a nonmagnetic singlet ground state (J = 0). However, recent theoretical and experimental studies suggest that novel magnetic states in the iridates with pentavalent $Ir^{5+}$($5d^4$) ions can also emerge from competitions between exchange interactions (0.05-0.10 eV) **[12],** non-cubic crystal fields, singlet-triplet splitting (0.050 - 0.20 eV) and SOI **[17-19]**.

Moreover, the vast majority of iridates studied thus far are either two- or three-dimensional systems **[4]**. Little work has been done on quasi-one-dimensional iridates, particularly those with dimers commonly found in other transition metal materials, and those in which the average number of electrons per transition metal is nonintegral, where double exchange is expected to play a critical role **[20]**. Double exchange occurs when one "extra" electron hops between neighboring transition metal ions with localized spins; and the Hund's rule $J_H$ coupling stabilizes the state with maximum possible (parallel) spin. This idea was first proposed by Zener for Mn dimers with average valence $Mn^{3.5+}$ **[21]**, and has been applied successfully to many materials, such as colossal magnetoresistive manganites **[22].** However, the strong SOI and covalency present in spin-dimer systems based on 5d-electrons are expected to compete with the Hund's rule coupling and double exchange, leading to new quantum phenomena. We show that such is indeed the case in $Ba_5AlIr_2O_{11}$.

In this paper, we report a novel magnetic state and dielectric behavior dictated by the joint action of a strong SOI and charge ordering in the dimer-chain of $Ba_5AlIr_2O_{11}$. This newly-synthesized, single-crystal iridate features both tetravalent $Ir^{4+}$ and pentavalent $Ir^{5+}$ ions that occupy dimers linked by $AlO_4$-tetrahedra lying along the **b-**axis (see **Fig. 1**). Despite their one-dimensional character, the dimer chains undergo a second-order structural change or charge order at $T_S = 210$ K, and a transition to magnetic order at $T_M =$ 4.5 K. The antiferromagnetic state below $T_M$ is highly anisotropic and resilient to strong magnetic field (up to 14 T), but is susceptible to even modest hydrostatic pressure (up to 10 kbar). We propose that the charge order consists of an ordered arrangement of $Ir^{4+}$ and $Ir^{5+}$ ions within each dimer, which forms electrical dipoles and promotes strong dielectric response that develops near $T_S$. Our studies suggest that a combined effect of both a strong SOI and covalent bonding overcomes double exchange and stabilizes coexisting antiferromagnetic and charge orders. The physical behavior of the dimer chain $Ba_5AlIr_2O_{11}$ provides additional evidence of new physics unique to materials with strong SOI.

*Experimental details* Single-crystals of $Ba_5AlIr_2O_{11}$ were synthesized using the self-flux method that is described elsewhere **[4]**. The average size of the single crystals is 2.0 x 1 x 1 $mm^3$ (**Fig. 1**). The crystal structures were determined using both a Nonius Kappa CCD X-Ray single-crystal diffractometer at the University of Kentucky, and a Rigaku X-ray diffractometer equipped with a PILATUS 200K hybrid pixel array detector at Oak Ridge National Laboratory. Full data sets were collected between 100K and 300K, and the structures were refined using FullProf software **[23].** Chemical compositions of the single crystals were estimated using both single-crystal X-ray diffraction and energy dispersive X-ray analysis (Hitachi/Oxford 3000). Magnetization, specific heat and electrical resistivity were measured using either a Quantum Design MPMS-7 SQUID Magnetometer and/or Physical Property Measurement System with 14-T field capability. The complex permittivity $\varepsilon(T,H,\omega) = \varepsilon' + i\varepsilon''$ was measured using a 7600 QuadTech LCR Meter capable of operating over the frequency range 10 Hz $\leq f \leq$ 2 MHz. The high-temperature resistivity was measured using a Displex closed-cycle cryostat capable of continuous temperature ramping from 9 K to 900 K. Pressure measurements were performed up to 1.3 GPa in the MPMS using a BeCu pressure cell and Delphi oil as the pressure transmitting medium.

*Crystal structure determination* $Ba_5AlIr_2O_{11}$ adopts an orthorhombic structure with space group Pnma (No. 62) as shown in **Fig. 1**. This newly synthesized single-crystal iridate features a non-integral valence state based on tetravalent $Ir^{4+}$ and pentavalent $Ir^{5+}$ states assigned to the iridate dimers that are linked by $AlO_4$-tetrahedra lying along the **b-**axis. The lattice parameters are a = 18.7630(38) Å, b = 5.7552(12) Å and c = 11.0649(22) Å at room temperature; they undergo a subtle change near $T_S$ = 210 K (**Figs. 2a** and **2b**). This change is much more pronounced in the Ir1-Ir2 distance and thermal displacement U in a dimer, as shown in **Figs. 2c** and **2d**, respectively. Each dimer consists of two face-sharing $IrO_6$-octohedra with two inequivalent, octahedral Ir2 and Ir1 sites occupied by tetravalent $Ir^{4+}(5d^5)$ and pentavalent $Ir^{5+}(5d^4)$ ions, respectively (see **Fig. 1b**). The average Ir1-O bond distance $d_{[Ir1-O]}$ = 1.997 Å, the average Ir2-O bond distance $d_{[Ir2-O]}$ = 2.013 Å, and the average Ir-Ir distance $d_{[Ir-Ir]}$ = 2.7204(5) Å at T=100 K that undergoes a sharp slope change at $T_S$ = 210 K (**Fig. 2c**). The fact that $d_{[Ir2-O]} > d_{[Ir1-O]}$ indicates that different oxidation states exist on the two Ir sites, i.e. there occurs charge ordering in this system. The longer $d_{[Ir2-O]}$ is most likely due to the relatively large ionic radius *r* of $Ir^{4+}$ (*r* = 0.625 Å and 0.570Å for $Ir^{4+}$ and $Ir^{5+}$, respectively), and is assigned to the $IrO_6$ octahedra that are corner-connected with the $AlO_4$-tetrahedra (**Fig. 1b**). The results of the band structure calculations justify the existence of the charge ordering **[24]**, although it is not a complete one, which is common in transition metal oxides **[20]**. A certain degree of order among the $Ir^{4+}$and $Ir^{5+}$ ions in each dimer may already exist at room temperature; however, the anomalies in the lattice parameters (**Fig.2**), electrical resistivity and dielectric constant (**Fig.3**), and specific heat observed at $T_S$ = 210 K (**Fig.6**) signal an enhanced degree of order among the $Ir^{4+}$and $Ir^{5+}$ ions -- i.e., the formation of charge order. All Ir-O dimers are corner-connected through $AlO_4$ tetrahedra, forming dimer chains along the **b-**axis, but the dimer-chains are not connected along the **a-** and **c-**axes (see **Figs. 1a** and **2b)**. This peculiar structural characteristic generates weak intra-chain interactions via long Ir-O-O-Ir pathways, and very small inter-chain interactions due to the lack of pathways between chains, which makes the observed long-range orders particularly unusual. All six Ir-O bond lengths in each octahedron are unequal, but the non-cubic crystal field generated by these distortions is not a significant perturbation when compared to the SOI, as discussed below.

*Results and discussion* We first argue for the existence of charge order, which has important implications for the physical properties of $Ba_5AlIr_2O_{11}$. The electrical resistivity $\rho_b$ along the dimer chain direction increases by nearly nine orders of magnitude when temperature is lowered from 750 K ($10^2$ Ω cm) to 80 K ($10^{11}$ Ω cm) (**Fig. 3a**). More importantly, $\rho_b$ exhibits a distinct slope change near $T_S = 210$ K, and follows an activation law reasonably well (better than power laws) in a temperature range of 200-750 K, which yields an activation energy gap $\Delta_a \approx 0.57$ eV (Inset in **Fig. 3a**). The more rapid increase in $\rho_b$ below $T_S = 210$ K indicates the charge order transition is accompanied by increased localization of electronic states, as shown in **Fig. 3a**. The dielectric constant ε(T) and specific heat C(T) (discussed below) are also consistent with a bulk transition to long-range order at $T_S$.

The charge-ordered state we envision inevitably leads to formation of an electric dipole in each dimer. These electrical dipoles are parallel to each other *within* each dimer-chain, but are expected to orient anti-parallel *between* dimer-chains in order to minimize electrical energy, as schematically illustrated in **Fig. 4** (red arrows). This also explains the unusually large peak in dielectric response near $T_S = 210$ K, as shown in **Fig. 3b**; the real parts $\varepsilon_a'(T)$ and $\varepsilon_b'(T)$ of the **a-**axis and **b-**axis dielectric constants, respectively, rise by two orders of magnitude and peak near $T_S$ due to lattice softening, an inevitable consequence of the structural phase transition. The observed strong peak in both $\varepsilon_a'(T)$ and $\varepsilon_b'(T)$ is stronger than that of well-known ferroelectrics such as $BaMnF_4$, $BiMnO_3$, $HoMnO_3$ and $YMnO_3$ **[25-27]**. In addition, two weaker anomalies occur near 130 K and 30 K, respectively (Inset in **Fig. 3b**). The rapid decrease in $\varepsilon_b'(T)$ below 30 K implies the lattice stiffens as long-range magnetic order is approached at $T_N = 4.5$ K, as discussed below. Both $\varepsilon_a'(T)$ and $\varepsilon_b'(T)$ exhibit strong frequency dependences that can signal relaxor behavior.

Given the quasi-one-dimensional nature of the crystal structure of $Ba_5AlIr_2O_{11}$, it is reasonable that three-dimensional correlations necessary for magnetic order are established at the rather low temperature, $T_M = 4.5$ K. The **b-**axis magnetization $M_b$ exhibits a peak at $T_M$, whereas the **a-** and **c-**axis magnetizations $M_a$ and $M_c$ rise below $T_M$ (**Fig. 5a**). The observed large magnetic anisotropy indicates that the SOI is significantly stronger than any possible non-cubic crystal field due to the distortions in dimers, which would alter the effect of SOI, resulting in more isotropic magnetic behavior. We propose that the magnetic moments are aligned ferromagnetically within each dimer chain, but antiferromagnetically between dimer chains to minimize magnetic dipole energy (see **Fig. 4,** black arrows). However, it is curious that the magnetic susceptibility $\chi_b$ along the **b-**axis systematically decreases with applied field below $T_M$ (see **Fig. 5b**) indicating enhanced antiferromagnetic compensation or reduced moment fluctuations along the dimer chain direction. Moreover, the magnetic transition is barely suppressed in strong magnetic fields up to 12 T (**Fig. 5b**), which contrasts with the conventional expectation that a 12-T magnetic field should be strong enough to completely suppress $T_M$ (= 4.5 K at zero field). Nevertheless, $T_M$ is susceptible to hydrostatic pressure, and shifts from 4.5 K at ambient pressure to 5 K at 8.2 kbar at the substantial rate of 0.61 K/GPa (see **Fig. 5c**), implying an enhanced magnetic interaction between the 5d electrons and a magnetoelastic effect. The weak field dependence coupled with the substantial pressure dependence of $T_M$ constitutes noteworthy characteristics of this dimer chain system.

Data fits to the Curie-Weiss law for $50 < T < 320$ K yield a Curie-Weiss temperature $\theta_{CW}$ = -14 K and effective moment $\mu_{eff}$ = 1.04 $\mu_B$/dimer, much smaller than the value 3.88 $\mu_B$ expected for an S = 3/2 system (see **Fig. 5a)**. The negative $\theta_{CW}$ implies antiferromagnetic coupling, whereas the reduced $\mu_{eff}$ results from the joint effect of the SOI and electron hopping between the two Ir1 and Ir2 sites **[24]**. The onset of long-range magnetic order is also corroborated by a sharp λ-peak in the specific heat C(T) at $T_M$ = 4.5 K, measured both at $\mu_o H$ = 0 and 9 T (see **Fig. 6a**), which is consistent with the magnetization behavior shown in **Fig. 5b**. An analysis of the C(T) data yields an entropy removal below $T_M$ of approximately 1.00 J/mole K, which is well below the value 11.37 J/mole K anticipated for a S = 3/2 system (**Fig. 6a**). A portion of the magnetic entropy may be removed at higher temperatures by the transition at $T_S$ = 210 K. Indeed, we observe a weak, but well-defined anomaly in C(T) near 210 K, which decreases entropy, as shown in **Fig. 6b**. This observation reinforces the remarkable trend that a very small entropy is removed at a well-defined magnetic transition in iridates studied so far **[4, 15, 16, 19]**.

Our ab-initio calculations with the generalized gradient approximation (GGA) (to be published elsewhere **[24]**) indicate that charge order is slightly more energetically favorable than the double exchange state, although the charge order is incomplete (the charge disproportionation $\delta n_{Ir1/Ir2}$ is approximately 0.3 electrons). This is consistent with an exotic ground state in which antiferromagnetic and charge order coexist. The calculations (including GGA + SOI) result in a total effective moment $\mu_{eff}$~1.04 $\mu_B$/dimer (compared to 2.0 $\mu_B$/dimer obtained in the GGA calculations without SOI), which is remarkably consistent with the experimental value discussed above, and supports the importance of SOI. The combined effect of both SOI and electron hopping apparently alters the delicate balance between the competing energies, and as a result, weakens the Hund's rule coupling $J_H$ that tends to maximize the spin moment.

*Conclusions* We have observed the coexisting charge and antiferromagnetic orders in a quasi-one-dimensional dimer system, $Ba_5AlIr_2O_{11}$. All our observations are consistent with a subtle ground state that is stabilized by an unusual interplay between charge disproportionation (due to the crystal structure), formation of molecular orbitals (due to strong covalency), a double-exchange mechanism (comparable $J_H$) and SOI.

While charge order is clearly manifested by the structural transition and anomalies in, ρ(T) and C(T) at $T_S$ = 210 K (**Figs. 2, 3** and **6b)**. The charge disproportionation of ~ 0.3 electrons is not fully developed, but is sufficient to induce the strong anomalies observed in the dielectric constant (**Fig. 3**). The magnetic behavior at $T_M$ = 4.5 K indicates an exotic, long-range order that is sensitive to pressure, but not magnetic fields (**Fig. 5**). It is plausible that the magnetic moments are aligned (along the

**b-**axis) ferromagnetically within each dimer chain, but antiferromagnetically between dimer chains, to minimize magnetic dipole energy (**Fig. 4**).

It is clear that a purely ionic model with strong SOI, which would support a J = 1/2 state in $Ir^{4+}$ ($5d^5$) ions and J = 0 state in $Ir^{5+}$ ($5d^4$) ions, is not entirely applicable here. The reduced total moment per dimer (1.04 $\mu_B$/dimer) is evidence of depressed double exchange. These observations suggest covalency plays a role in formation of singlet molecular orbitals among some of the d-orbitals **[24, 28-32]**, but it alone cannot suppress Hund'e rule coupling $J_H$ or double exchange. Indeed, both covalency and the SOI conspire to suppress the double exchange in Ir dimers and stabilize a magnetic state that lies "in between" S =3/2 or J =1/2. Thus the combined action of strong covalency and the SOI overcomes DE and stabilizes a novel magnetic state.

The dimer chain $Ba_5AlIr_2O_{11}$ provides a unique paradigm for the investigation of SOI in the 4d/5d transition metal oxides, and especially mixed-valent 4d and 5d systems.

*Acknowledgements* GC is thankful to Drs. Rihbu Kaul and George Jackeli for useful discussions. This work was supported by the National Science Foundation via Grant No. DMR-1265162.

*Email: cao@uky.edu

**References**

1. B. J. Kim, Hosub Jin, S. J. Moon, J.-Y. Kim, B.-G. Park, C. S. Leem, Jaejun Yu, T. W. Noh, C. Kim, S.-J. Oh, V. Durairai, G. Cao & J.-H. Park *Phys. Rev. Lett*. 101, 076402 (2008)
2. S.J. Moon, H. Jin, K.W. Kim, W.S. Choi, Y.S. Lee, J. Yu, G. Cao, A. Sumi, H. Funakubo, C. Bernhard, and T.W. Noh *Phys. Rev. Lett*. 101, 226401 (2008)
3. B. J. Kim, H. Ohsumi, T. Komesu, S. Sakai, T. Morita, H. Takagi and T. Arima, *Science* **323**, 1329 (2009)
4. Gang Cao and Lance E DeLong, *"Frontiers of 4d- and 5d- Transition Metal Oxides"*, Singapore, *World Scientific* ISBN: 978-981-4374-859, June 2013
5. G. Cao, J. Bolivar, S. McCall, J.E. Crow, and R.P. Guertin, *Phys. Rev. B*, **57**, **R** 11039 (1998)
6. G. Cao, J.E. Crow, R.P. Guertin, P. Henning, C.C. Homes, M. Strongin, D.N. Basov, and E. Lochner, *Solid State Comm.* **113**, 657 (2000)
7. G. Cao, Y. Xin, C. S. Alexander, J.E. Crow and P. Schlottmann, *Phys. Rev. B* **66**, 214412 (2002)
8. G. Cao, V. Durairaj, S. Chikara, S. Parkin and P. Schlottmann, *Phys. Rev. B* 75, 134402 (2007)
9. G. Cao, S. Chikara, X.N. Lin, E. Elhami and V. Durairaj, *Phys. Rev. B* **69**, 174418 (2004)
10. O. B. Korneta, S. Chikara, L.E. DeLong. P. Schlottmann and G. Cao, *Phys. Rev. B* 81, 045101 (2010)

11. G. Cao, T. F. Qi, L. Li, J. Terzic, V. S. Cao, S. J. Yuan, M. Tovar, G. Murthy, and R. K. Kaul, *Phys. Rev B* **88** *220414 (R)* (2013)
12. Feng Ye, Songxue Chi, Huibo Cao, Bryan Chakoumakos, Jaime A. Fernandez-Baca, Radu Custelcean, Tongfei Qi, O. B. Korneta, and G. Cao, *Phys. Rev. B* **85**, 180403(R) (2012)
13. Feng Ye, Songxue Chi, Bryan C. Chakoumakos, Jaime A. Fernandez-Baca, Tongfei Qi, and G. Cao, *Phys. Rev. B* **87***,* 140406(R) (2013)
14. D. Haskel, G. Fabbris, Mikhail Zhernenkov, M. van Veenendaal, P. Kong, C. Jin, G. Cao**,** *Phys. Rev Lett.* **109**, 027204 (2012)
15. S. Chikara, O. Korneta W. P. Crummett, L. E. DeLong, P. Schlottmann and G. Cao, Phys. Rev. B **80** 140407 **(**R) **(**2009)
16. L. Li, P. P. Kong, T. F. Qi, C. Q. Jin, S. J. Yuan, L. E. DeLong, P. Schlottmann and G. Cao, *Phys. Rev. B* **87**, 235127 (2013)
17. Giniyat Khaliullin, Phys. Rev. Lett. 111, 197201 (2013)
18. G. Chen, L. Balents, and A. P. Schnyder, Phys. Rev. Lett. 102 096406 (2009)
19. G. Cao, T. F. Qi, L. Li, J. Terzic, S. J. Yuan, L. E. DeLong, G. Murthy and R. K. Kaul, in *Phys. Rev. Lett.* **112**, 056402 (2014)
20. D.I. Khomskii, Transition Metal Compounds, Cambridge University Press 2014
21. C. Zener, *Phys. Rev.* **82**, 403 (1951)
22. Yurii A Izyumov and Yu N Skryabin 2001 *Phys.-Usp.* **44** 109
23. J. Rodriguez-Carvajal, Physica B. **192,** 55 (1993)
24. S.V. Streltsov, J. Terzic, J. C. Wang, Feng Ye, , W. H. Song, S. J. Yuan, S. Aswartham, and D.I. Khomskii and G. Cao, to be submitted to Phys. Rev. Lett.,

2015

25. J. F. Scott, Phys. Rev. B **16**, 2329 (1977)
26. T. Kimura, S. Kawamoto, I. Yamada, M. Azuma, M. Takano and Y. Tokura, Phys. Rev B **67**, 180401(R) (2003)
27. B. Lorenz, Y. Q. Wang, Y. Y. Sun, and C. W. Chu, Phys. Rev. B **70**, 212412 (2004)
28. Sergey V. Streltsov and Daniel I. Khomskii, Phys. Rev. B **89**, 161112(R) (2014)
29. V. I. Anisimov, I. A. Nekrasov, D. E. Kondakov, T. M. Rice, and M. Sigrist, Eur. Phys. J. B **25**, 191 (2002).
30. A. Koga, N. Kawakami, T. Rice, and M. Sigrist, Phys. Rev. Lett. **92**, 216402 (2004).
31. L. De'Medici, A. Georges, and S. Biermann, Phys. Rev. B - Condens. Matter Mater. Phys. **72**, 205124 (2005)
32. S. V Streltsov and D. I. Khomskii, Phys. Rev. B **89**, 161112 (2014)

**Captions:**

**Fig.1.** Crystal structure: The single-crystal structure generated based on the x-ray data for **(a)** the ac-plane, **(b)** the ab-plane and **(c)** a representative single crystal of $Ba_5AlIr_2O_{11}$. Note that the dimers are connected via $AlO_4$-tetrahedra (yellow), forming dimer-chains along the **b**-axis.

**Fig.2.** The temperature dependence of the lattice parameters of single-crystal $Ba_5AlIr_2O_{11}$ **(a)** the **a**-axis, **b**-axis, and **c**-axis (A = a-, b- or c-axis; $A_{300}$ = the lattice parameter at 300 K) **(b)** the unit cell V, **(c)** the Ir1-Ir2 distance (S1=sample 1 and S2=sample 2) and **(d)** the thermal displacement U for 90 K < T < 300 K. Note the pronounced changes at $T_S$=210 K in the Ir1-Ir2 distance and thermal displacement U.

**Fig.3.** Temperature dependence of **(a)** the **b**-axis electrical resistivity $\rho_b$ and **(b)** the dielectric constant for the **a**-axis and **b**-axis $\varepsilon_a$' and $\varepsilon_b$', respectively. Inset in **(a)**: ln $\rho_b$ vs. 1/T; Inset in **(b)**: $\varepsilon_b$' vs. T for lower temperatures.

**Fig. 4.** A sketch for the proposed configurations of electrical dipoles (E, red arrows) and magnetic moments (M, black arrows) for the ac-plane (left) and ab-plane (right) based on the data collected for this study

**Fig. 5.** Temperature dependence of **(a)** the magnetization for the **a**-, **b**- and **c**-axis, $M_a$, $M_b$ and $M_c$ at $\mu_oH$=7 T, **(b)** the b-axis magnetic susceptibility $\chi_b$ at various fields, and **(c)** $M_b$ at various pressure and $\mu_oH$=7 T.

**Fig. 6.** Temperature dependence of **(a)** the specific heat C(T) at $\mu_oH$=0 and 9 T and entropy at $\mu_oH$=T (right scale) and **(b)** C(T) for 140 K < T < 300 K.

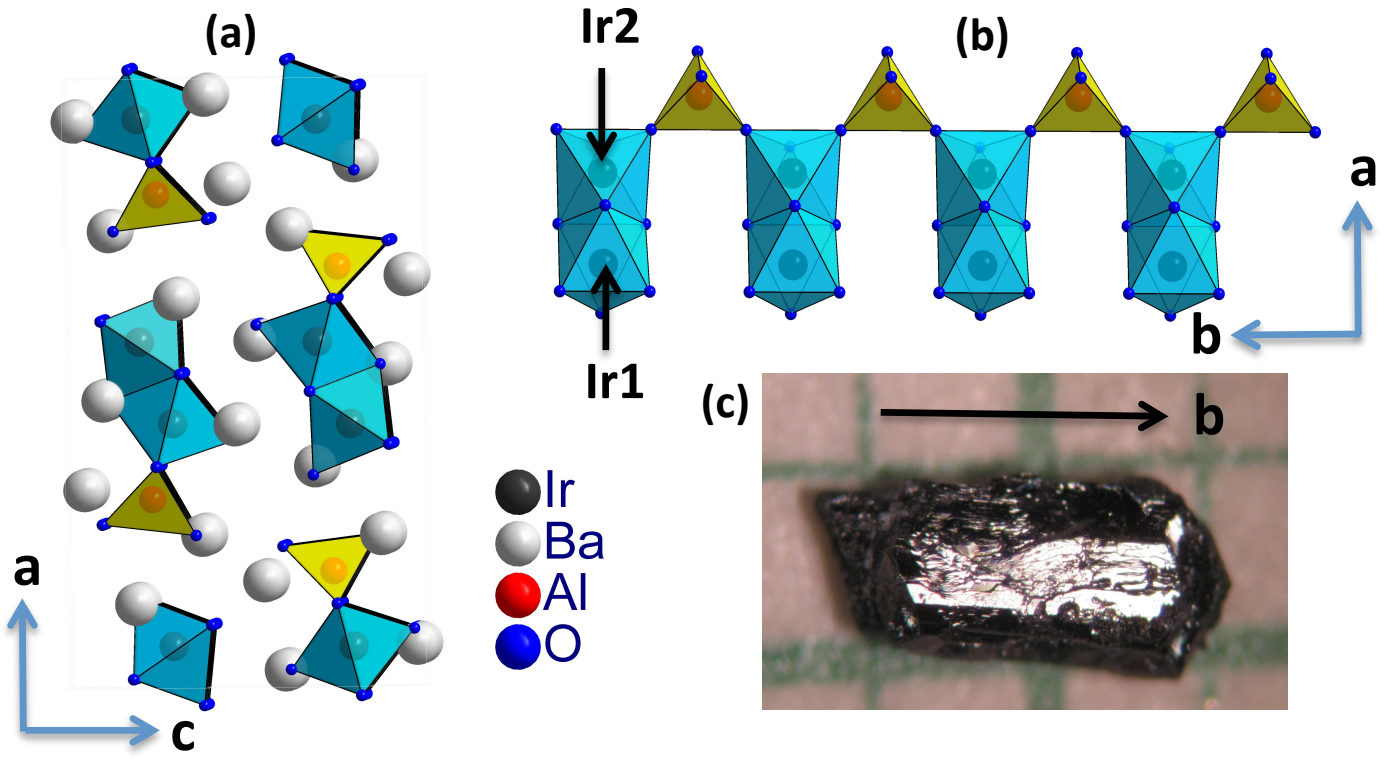

Fig. 1

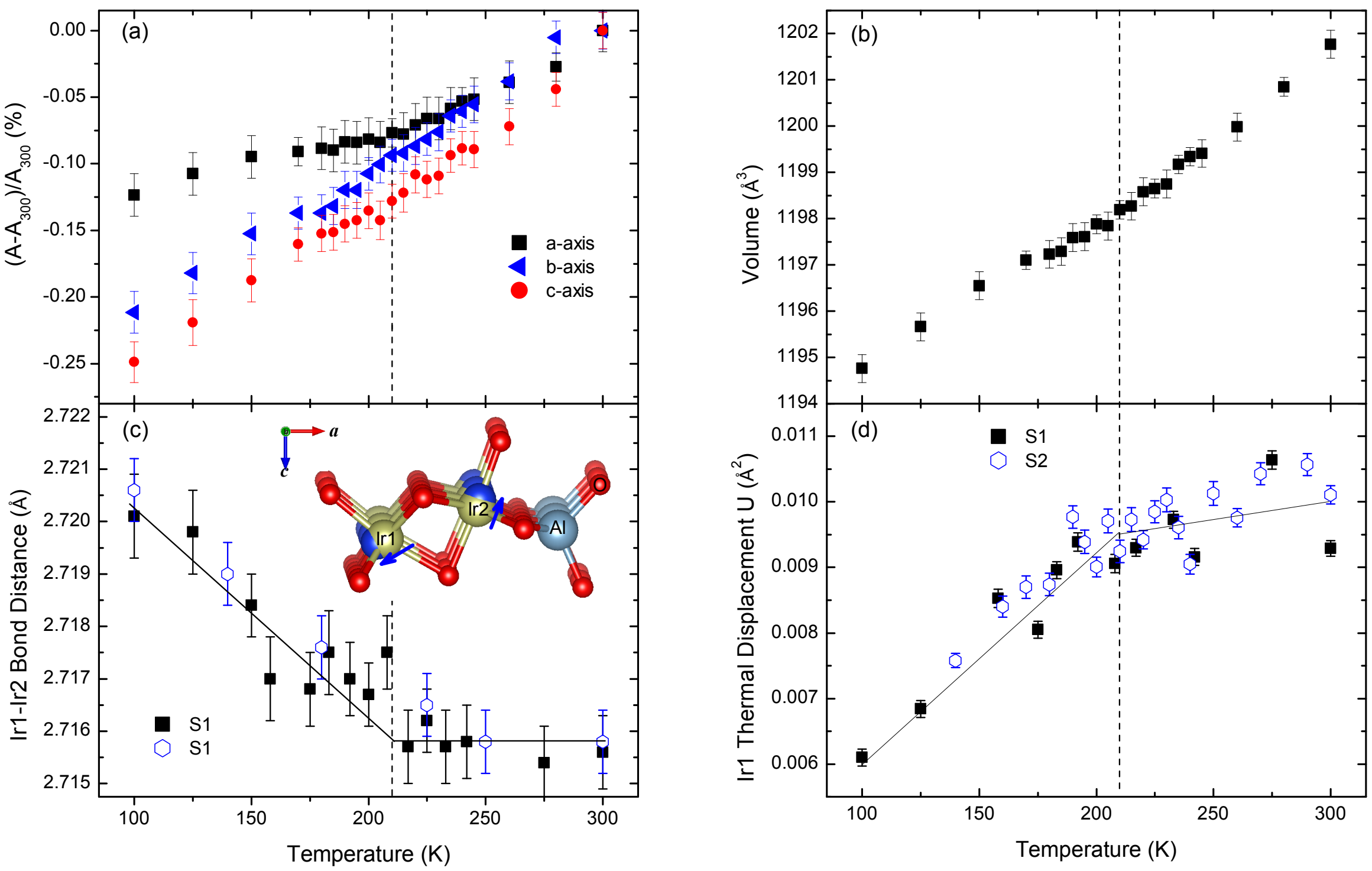

Fig. 2

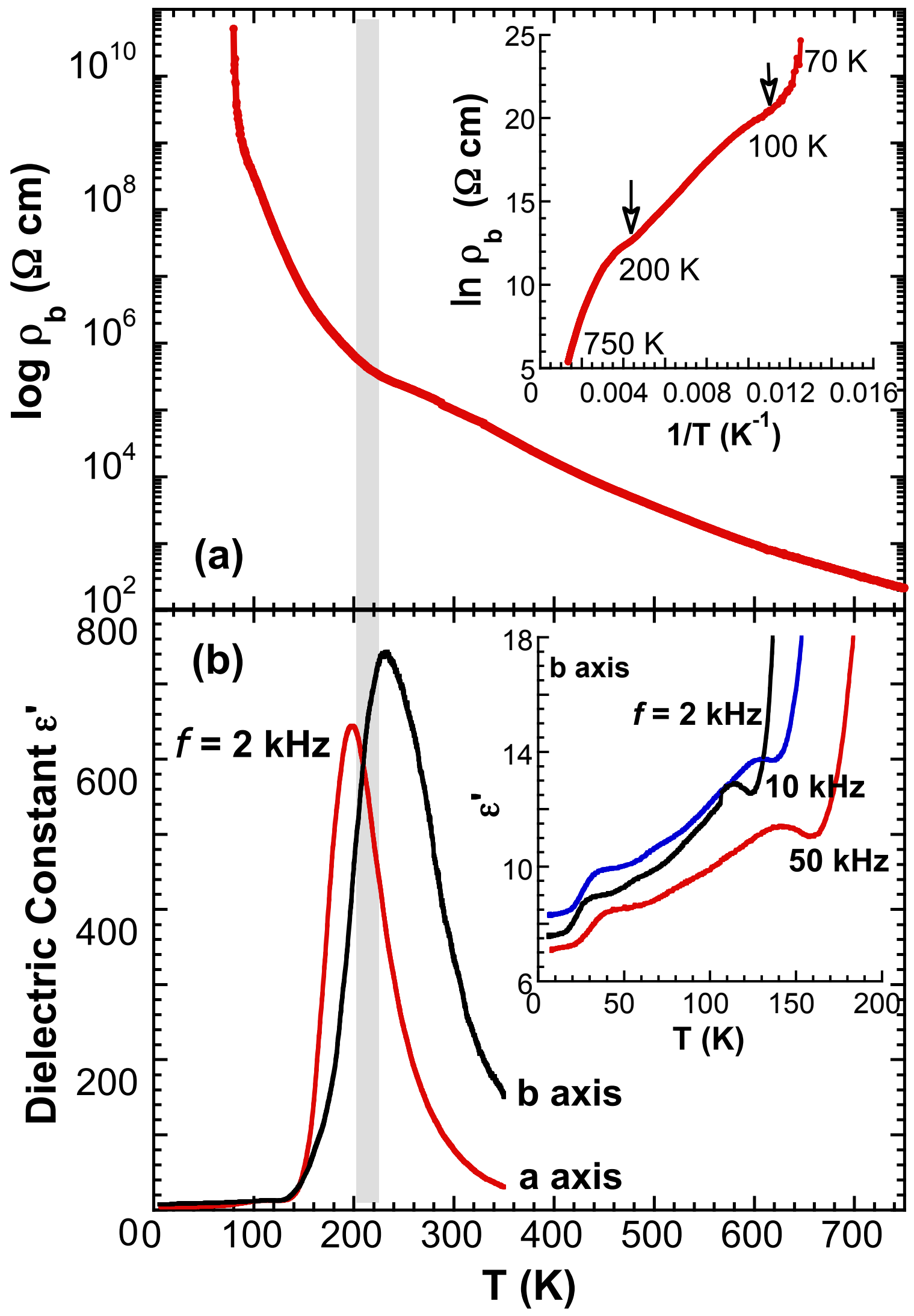

Fig. 3

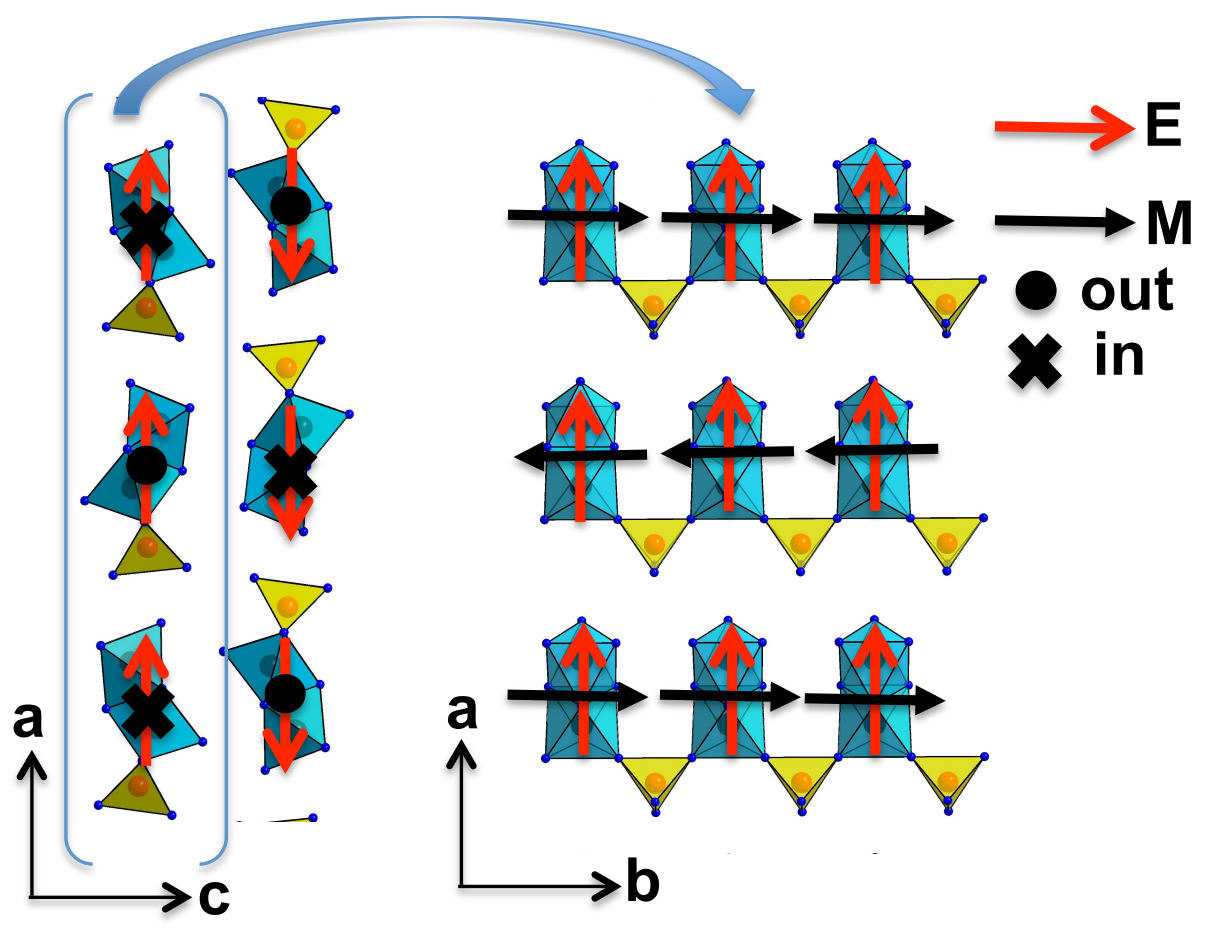

Fig. 4

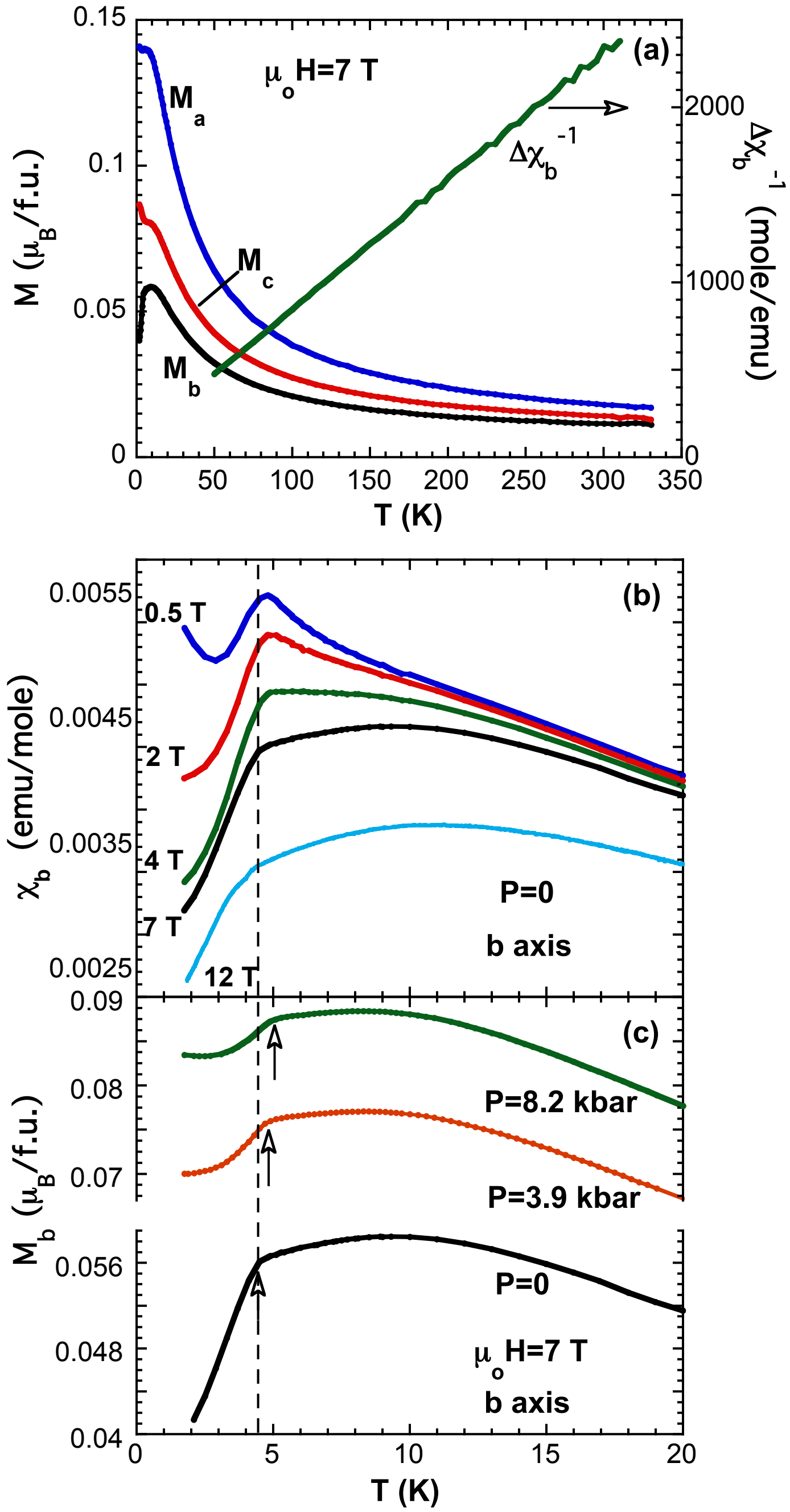

Fig. 5

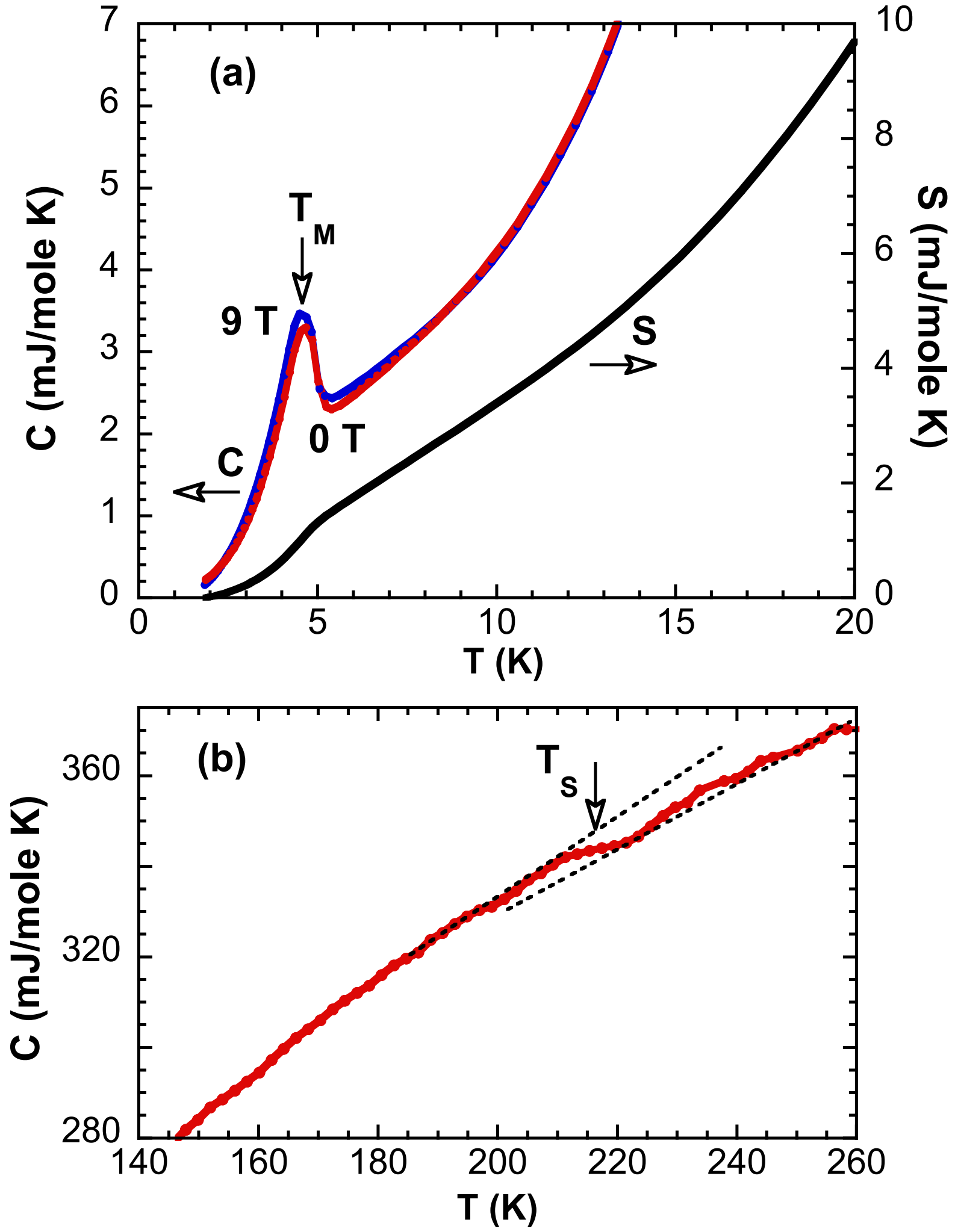

Fig. 6